\documentclass[%
 reprint,
 amsmath,amssymb,
 aps,
]{revtex4-2}
\usepackage{graphicx}
\usepackage{dcolumn}
\usepackage{bm}
\usepackage{xcolor}
\usepackage{microtype}
\usepackage{flushend}

\begin{document}

\preprint{APS/123-QED}

\title{Is the active suspension in a complex viscoelastic fluid more chaotic or more ordered?}

\author{Yuan Zhou$^{1,2}$}

\author{Qingzhi Zou$^{1}$}

\author{Ignacio Pagonabarraga$^{3,4}$}
\email{Contact author: ipagonabarraga@ub.edu}

\author{Kaihuan Zhang$^{1,2}$}

\author{Kai Qi$^{1,2}$}
\email{Contact author: kqi@mail.sim.ac.cn}

\affiliation{
$^{1}$2020 X-Lab, Shanghai Institute of Microsystem and Information Technology, Chinese Academy of Sciences, Shanghai 200050, China \\
$^{2}$Center of Materials Science and Optoelectronics Engineering, University of Chinese Academy of Sciences, Beijing 100049, China \\
$^{3}$Departament de F\'isica de la Mat\`eria Condensada, Universitat de Barcelona, C. Mart\'i Franqu\`es 1, 08028 Barcelona, Spain \\
$^{4}$University of Barcelona Institute of Complex Systems (UBICS), Universitat de Barcelona, 08028 Barcelona, Spain
}

\begin{abstract}
The habitat of microorganisms is typically complex and viscoelastic. A natural question arises: Do polymers in a suspension of active swimmers enhance chaotic motion or promote orientational order? We address this issue by performing lattice Boltzmann simulations of squirmer suspensions in polymer solutions. At intermediate swimmer volume fractions, comparing to the Newtonian counterpart, polymers enhance polarization by up to a factor of 26 for neutral squirmers and 5 for pullers, thereby notably increasing orientational order. This effect arises from hydrodynamic feedback mechanism: squirmers stretch and align polymers, which in turn reinforce swimmer orientation and enhance polarization via hydrodynamic and steric interactions. The mechanism is validated by a positive correlation between polarization and a defined polymer–swimmer alignment parameter. Our findings establish a framework for understanding collective motion in complex fluids and suggest strategies for controlling active systems via polymer-mediated interactions.
\end{abstract}

\maketitle

Microswimmer locomotion in viscoelastic fluids plays a fundamental role in fertilization, infection, and biofilm formation \cite{Hyakutake2015sperm, Fauci2006biofluidmechanics, Montecucco2001helicobacter, Celli2009helicobacter, Wilking2011biofilms}. Complex fluids significantly alter swimming kinematics, motility, and transport properties, with effects strongly dependent on swimmer type and fluid rheology \cite{patteson2015running,Qin2015Flagellar,Zoettl2023Dynamics,Zoettl2019Enhanced,Riley2017Empirical,Gomez2016JanusParticle,Narinder2018Achiral,Li2015Undulatory,Qi2020Squirmer,qi2025unravel,qi2020rheotaxis,Corato2021Spontaneous,Corato2025Janusdisk,Bozorgi2011Collective, Bozorgi2014Nonlinear, Hemingway2015Active,Hemingway2016Viscoelastic,Bozorgi2011Collective,mo2023hydrodynamic,zhang2023locomotion,Tung2017sperm,gonzalez2025morphogenesis,liao2023viscoelasticity,Li2016Collective}. For instance, \textit{E. coli} swims faster and straighter in polymer solutions due to flagellar-induced polymer stretching, which generates elastic stresses that suppress flagellar unbundling \cite{patteson2015running}. In contrast, \textit{C. reinhardtii} experiences restricted flagellar motion near the cell body, leading to reduced swimming speed despite increased beating frequency \cite{Qin2015Flagellar}. Hydrodynamic simulations reveal that Bacterial swimming speeds can increase in polymeric fluids despite higher viscosity, driven by non-uniform polymer distribution and flagellar chirality \cite{Zoettl2019Enhanced}. Recent experiments demonstrate enhanced rotation motion of self-propelled Janus particle in viscoelastic fluids \cite{Gomez2016JanusParticle}, with enhanced diffusion transitioning to persistent circular motion above a critical Deborah number \cite{Narinder2018Achiral}. Coarse-grained simulations suggest that rotational enhancement is driven by polymer adsorption and asymmetric polymer encounters, with optimal effects near the overlap concentration \cite{Qi2020Squirmer,qi2025unravel}. Beyond individual locomotion, viscoelasticity plays a crucial role in collective behavior \cite{Bozorgi2011Collective, Bozorgi2014Nonlinear, Hemingway2015Active, Li2016Collective, Hemingway2016Viscoelastic,Tung2017sperm,mo2023hydrodynamic,zhang2023locomotion,gonzalez2025morphogenesis,liao2023viscoelasticity}. Bovine spermatozoa, which swim disordered in Newtonian fluids, can assemble into bundles \cite{mo2023hydrodynamic,zhang2023locomotion} and form dynamic clusters in viscoelastic media \cite{Tung2017sperm}, with alignment increasing as fluid elasticity rises. Bacteria can generically and reversibly assemble into large serpentine ``cables'' during proliferation, driven by the interplay between polymer-induced entropic attractions among neighboring cells and the reduced ability of the cells to diffuse apart in a viscous polymer solution \cite{gonzalez2025morphogenesis}. Furthermore, the length scale of collective bacterial locomotion increases monotonically with the suspension’s elasticity \cite{liao2023viscoelasticity}. Simulations of rod-like pushers in two-dimensional films further reveal that viscoelasticity enhances aggregation, generating strong polymer stresses in interstitial gaps between aligned swimmers \cite{Li2016Collective}. However, the role of polymer deformations in modulating collective swimmer behavior in bulk remains underexplored, presenting an open question in the study of active matter in complex fluids.

In this work, we perform hydrodynamic simulations of multiple squirmers in polymer solutions using the lattice Boltzmann method (LBM)~\cite{lbm2001, Nash2008lbm, Pozrikidis1992Boundary} to investigate their clustering behavior. The collective orientation, quantified by the polarization, increases by a factor of 26 for neutral squirmers and 5 for pullers at intermediate swimmer volume fractions relative to the polymer-free system. To unravel the underlining mechanism, the cluster-size distribution and polymer stretching parameters are analyzed, including the radius of gyration and anisotropy parameter. Our results indicate a hydrodynamic feedback mechanism in which squirmers stretch polymers, inducing their deformation and alignment with the swimming direction, which in turn reinforces swimmer orientation and amplifies polarization. This interpretation is supported by the coherent variation between a defined alignment parameter and the polarization, highlighting the role of polymer deformation in shaping the collective behavior of microorganisms.

\begin{figure}[t]
\includegraphics[width=0.85\columnwidth]{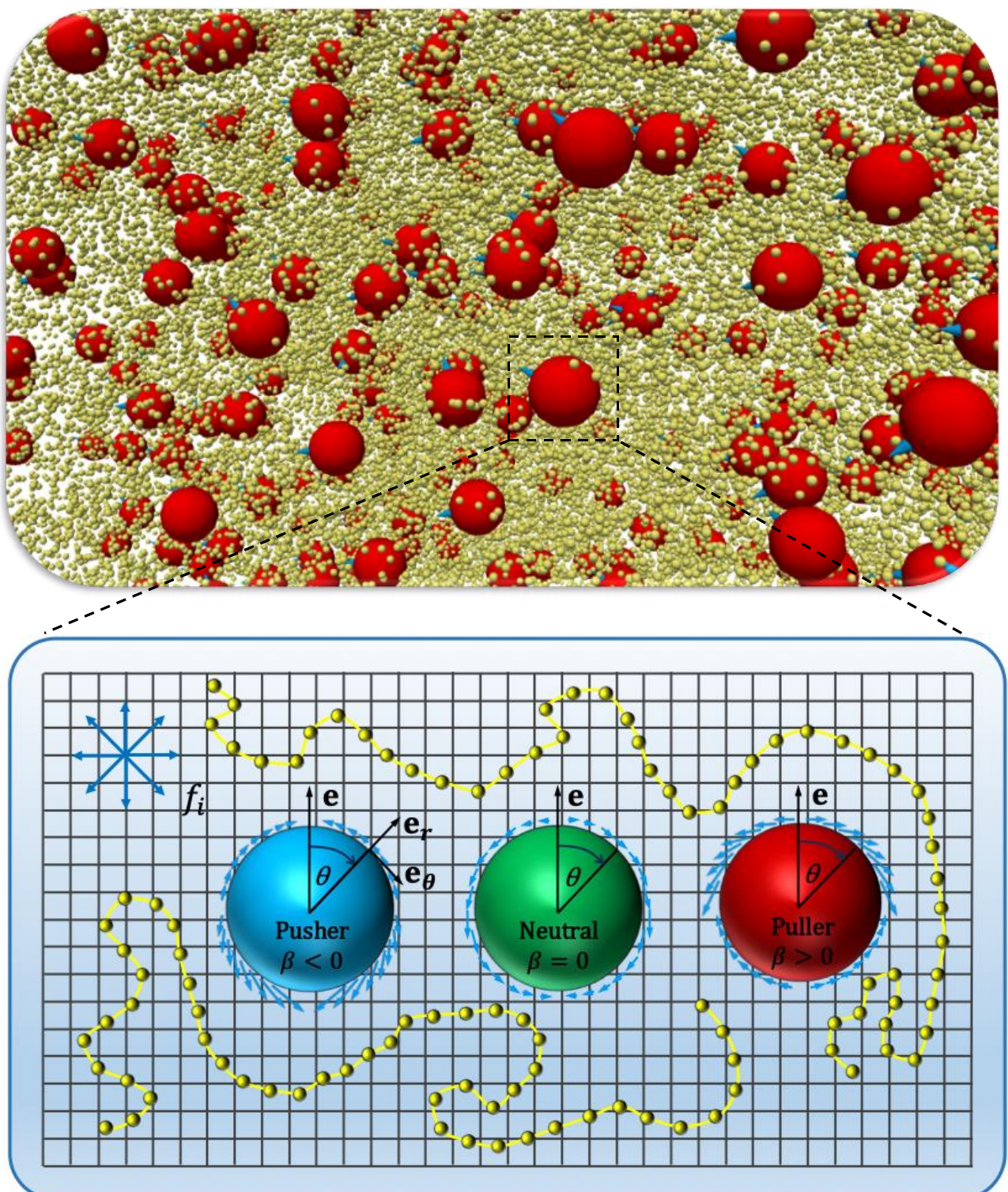}
    \caption{\label{fig:system} Simulation snapshot and schematic of a system containing multiple squirmers and polymers. In the snapshot, red spheres represent pullers, and yellow chains represent flexible polymers. Swimmer types---pushers, neutral squirmers, and pullers---are distinguished by their respective $\beta$ values. Hydrodynamics are resolved using the lattice Boltzmann method.}
\end{figure}

We model each microswimmer as a spherical squirmer with prescribed tangential surface slip velocity \cite{SquirmingMotion1952, SphericalEnvelopeApproach1971,ActiveParticle2017,Theers2016Microswimmer, Theers2018Clustering, Qi2022ActuveTurbulence}
\begin{equation}
    {\bf u}_s=B_1 \sin(\theta)[1+\beta \cos(\theta)] \mathbf{e}_{\theta}
\end{equation}
where $\theta = \arccos(\mathbf{e} \cdot \mathbf{e}_r)$ is the polar angle between the radial unit vector $\mathbf{e}_r$ and the swimming direction $\mathbf{e}$, and $\mathbf{e}_\theta$ denotes the local tangent vector. The swimming speed $U_0 = 2B_1/3$ is set by the first squirming mode $B_1$ of the Legendre polynomial expansion. The dimensionless parameter $\beta$ characterizes the active stress ($\beta < 0$ pusher, $\beta = 0$ neutral, $\beta > 0$ puller; see Fig.~\ref{fig:system}). The surrounding viscoelastic medium is modeled as an inertialess flexible polymer chain of natural bond length $l^n_0$ \cite{Nash2008lbm,qi2025unravel}, consisting of $N_m$ monomers connected via finitely extensible nonlinear elastic (FENE) bonds \cite{Kremer1990Entangled} (see Supplemental Material, SM). Additionally, soft-sphere interactions are employed to account for volume-exclusion forces between polymer-polymer, polymer-squirmer, and squirmer-squirmer pairs. Hydrodynamics are resolved using the LBM in a cubic box of size $L/l^n_0 = 80$ with periodic boundary conditions in all directions (Fig.~\ref{fig:system}). We consider squirmers of radius $R/l^n_0=3.0$, with their volume fraction given by $\phi_s = 4 \pi  R^3 N_s/3 L^3$, where $N_s$ denotes the number of squirmers. Similarly, the polymer concentration is defined as $\phi_p = 4\pi N_p R_{g0}^3/3L^3$, where $N_p$ represents the number of polymer chains, and $R_{g0}$ refers to the radius of polymer gyration in a dilute squirmer suspension (see SM). The simulations are conducted under athermal conditions.

\begin{figure}[!htbp]
\includegraphics[width=0.9\columnwidth]{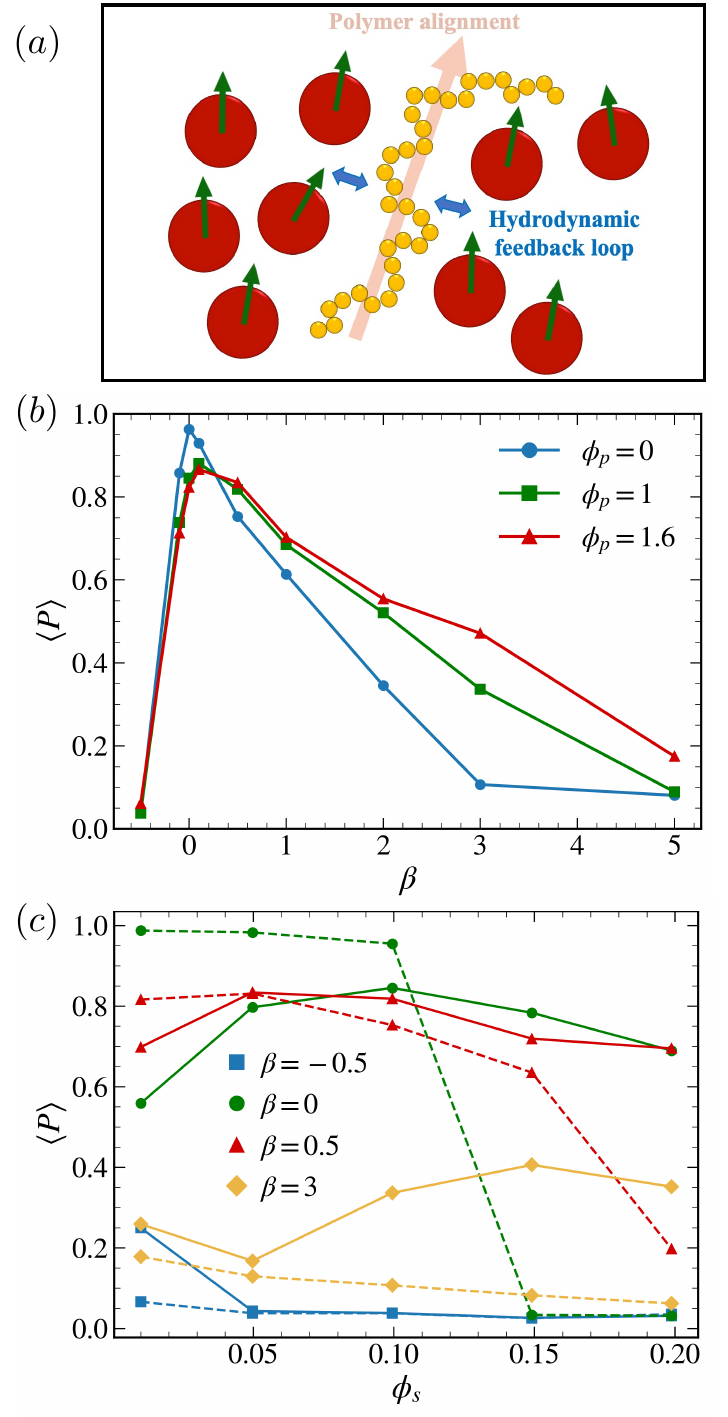}
\caption{\label{fig:polarization} 
    (a) Schematic illustration of the hydrodynamic feedback mechanism. (b) Dependence of polarization on active stress at various polymer concentrations, with swimmer concentration fixed at $\phi_s = 0.1$. (c) Dependence of polarization on swimmer concentration for different types of swimmers. Solid and dashed lines represent polymer concentrations $\phi_p = 1.0$ and $\phi_p = 0$, respectively.} 
\end{figure}

Previous studies \cite{polar2013} have shown that active stresses play a central role in shaping the collective behavior of squirmers. The degree of alignment is quantified by the global polar order parameter
\begin{equation}
    P(t)=\frac{\sum_{i=1}^{N_s}\mathbf{e}_i}{N_s}
\end{equation}
where $\mathbf{e}_i$ is the intrinsic orientation of the $i$-th squirmer. In Newtonian fluids (see Fig.~\ref{fig:polarization}(b)), neutral squirmers exhibit high polarization ($P \approx 0.96$), pullers show decreasing polarization as $\beta$ increases, and pushers maintain near-zero polarization. This behavior arises from the nature of the hydrodynamic interactions dictated by the active stress: neutral swimmers preserve nearly perfect order because the irrotational flow from their translational motion is sufficient to maintain a symmetric fluid distortion; pushers generate active stresses favoring head-to-head configurations that compete with self-propulsion, leading to rapid decorrelation; and pullers destabilize head-to-head configurations, promoting alignment along a common direction \cite{polar2013}. In polymeric fluids (see Fig.~\ref{fig:polarization}(b) and Fig.~S5(a)), polarization is enhanced by a factor of $1.1 \sim 4.7$ across a wide range of $\beta = 0.5 \sim 5$. As illustrated in Fig.~\ref{fig:polarization}(a), this striking effect originates from a hydrodynamic feedback loop in which strong swimmer-induced flows stretch polymers, reinforcing their alignments with the surrounding squirmers. In turn, the stretched polymers act as soft confinements, modulating the flow fields and further promoting polarization via hydrodynamic and steric interactions. For $\beta \approx -0.1 \sim 0.1$, the swimmer-generated flows are too weak to trigger this feedback loop. As a consequence, polymers instead distort rather than regularize the flows, thus suppressing the order. At high polymer concentration ($\phi_p = 1.6$), the feedback mechanism can be activated even for weak pushers, yielding a 1.7-fold increase in polarization (see Fig.~S5(a)). It is worthy noting that this mechanism stems from the long-chain nature of the polymer. Shorter polymers or monomer solutions would lead to a reduction or even the absence of polarization enhancement (see Fig.~S3). 

The interplay between polymer and squirmer concentrations on polarization is shown in Fig.~\ref{fig:polarization}(c) and Fig.~S5(b). For neutral swimmers in a Newtonian fluid, polarization decreases sharply with increasing squirmer concentration, dropping from $0.96$ at $\phi_s = 0.10$ to $0.03$ at $\phi_s = 0.15$, signaling a transition from hydrodynamically aligned motion to a disordered state. This loss of order is attributed to the increased flow-field variability at higher squirmer concentrations, hindering the synchronization of swimming directions. In polymer solutions, the source-dipole flow generated by dilute neutral swimmers is insufficient to stretch polymers. Therefore, their presence suppresses the system order, leading to a reduction of polarization by 10\% to 45\% for $\phi_s <0.1$. Remarkably, at higher squirmer concentrations, stronger flows generated by swimmers stretch and align the polymers, activating the hydrodynamic feedback loop and thus enhancing the polarization by at most 26 times.  

For pullers with $\beta = 0.5$, the polarization trend is similar to that of neutral swimmers. However, the stronger force-dipole flow generated by pullers shifts the transition from polarization suppression to enhancement to a smaller squirmer concentration, $\phi_s = 0.05$. At high concentration ($\phi_s = 0.20$), the hydrodynamic feedback mechanism yields a $3.45$-fold enhancement in polarization. 
For pullers with $\beta = 3$, the presence of polymers induces only a slight promotion of system order in dilute swimmer suspensions. As $\phi_s$ increases ($\phi_s \geq 0.05$), the substantially stronger flows cause significant polymer stretching, allowing polymer-induced alignment to remain effective over a broader range of concentrations and producing enhancements of up to a factor of $5.83$ at $\phi_s = 0.20$ (see Fig.~\ref{fig:polarization}(c) and Fig.~S5(b)).

For pushers, polarization is initially low at $\phi_s=0.01$ due to hydrodynamic interactions disrupting alignment. However, the introduction of polymers initiates the hydrodynamic feedback mechanism, leading to a polarization increase by a factor of 3.57 (see Fig.~\ref{fig:polarization}(c) and Fig.~S5(b)). Nevertheless, for $\phi_s \ge 0.05$, hydrodynamic interferences become dominant, causing alignment to diminish regardless of the presence of polymers.

\begin{figure*}[t]
\includegraphics[width=2.07\columnwidth]{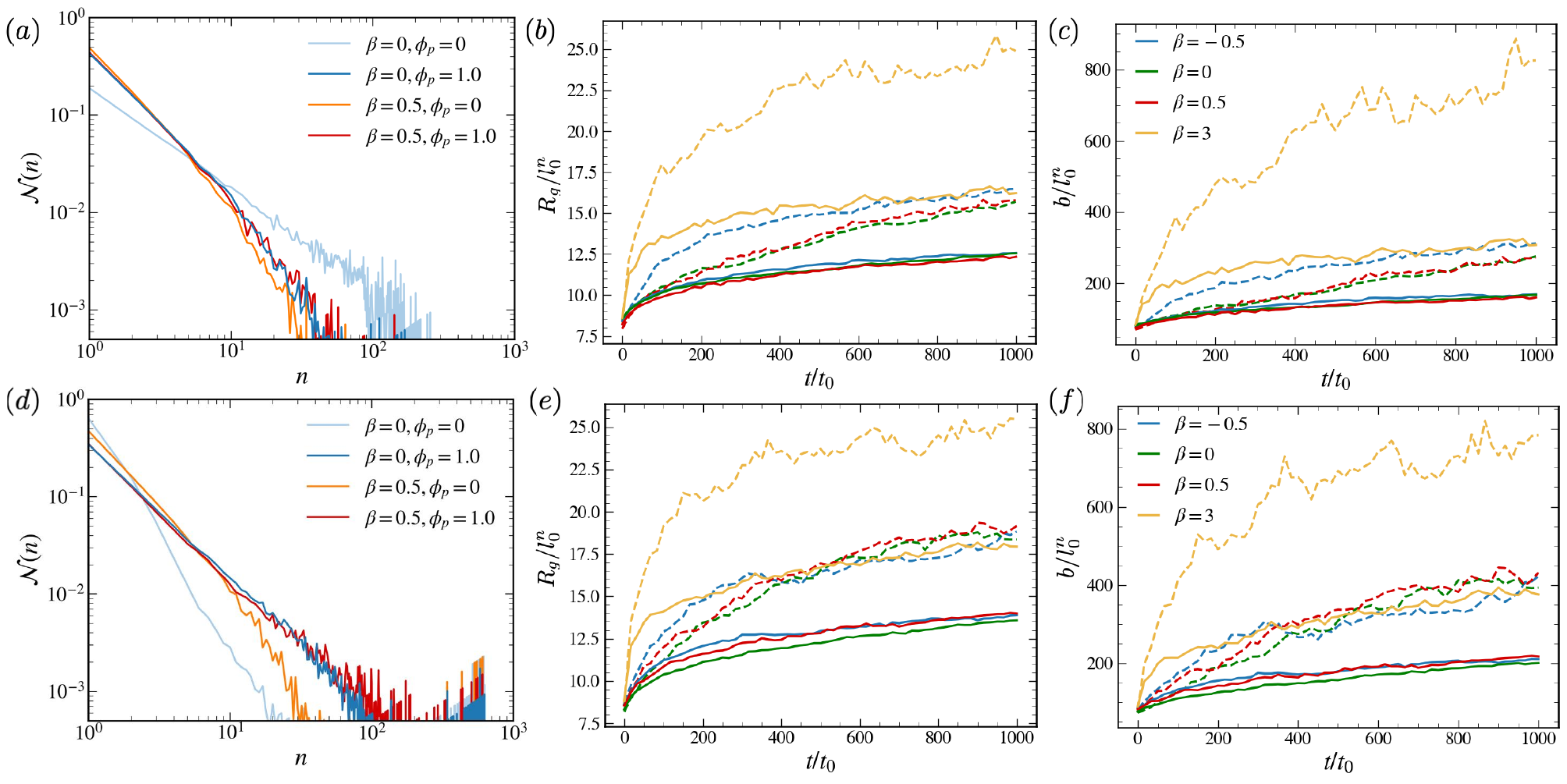}
    \caption{\label{fig:wide} (a-c) Cluster size distribution, radius of gyration, and anisotropy parameter $b$ at $\phi_s=0.1$.  (a) Cluster size distribution for neutral swimmers ($\beta=0$) and puller ($\beta=0.5$), in the presence and absence of polymers. (b) Radius of gyration for various active stresses, shown for both $\phi_p=1.0$ (solid lines) and $\phi_p=0.15$ (dashed lines) systems. $t_0=3r/B_1$ denotes the characteristic time. (c) Anisotropy parameter $b$ for different active stresses. (d-f) Same analyses conducted at $\phi_s=0.15$.}
\end{figure*}

To characterize the clustering behavior of squirmers, we evaluate the cluster-size distribution \cite{Clustering2014, Bacterial2020, Qi2022ActuveTurbulence} 
\begin{equation}
    \mathcal{N}(n)=\frac{n}{N_s}G(n), 
\end{equation} 
defined as the fraction of squirmers belonging to a cluster of size $n$, with formation probability $G(n)$. The distribution satisfies the normalization $\sum_{n=1} \mathcal{N}(n) = 1$. A cluster is defined based on both distance and orientation criteria: a squirmer is considered part of a cluster if its center-of-mass distance to another squirmer satisfies $d_c/l^n_0<6.11$ and the angle between their orientations is less than the threshold $\theta<\pi/ 6$. As shown in Fig~\ref{fig:wide}(a), at $\phi_s=0.1$, polymers inhibit the formation of large clusters while promoting small- and intermediate-sized clusters for neutral swimmers, consistent with the observed reduction in polarization. In contrast, polymers slightly enhance alignment for pullers, increasing the probability of forming larger clusters. At $\phi_s = 0.15$ (Fig.~\ref{fig:wide}(d)), due to the hydrodynamic feedback mechanism, polymers enhance alignment for both neutral swimmers and pullers, increasing the likelihood of large-cluster formation ($n=20 \sim 100$). Note that the impact of polymers on cluster sizes vanishes at very low polymer concentrations, e.g., $\phi_p=0.15$ (see Fig.~S6). For pushers and pullers with large $\beta$, strong hydrodynamic interactions limit cluster sizes ($n \leq 10$), although increasing $\phi_s$ promotes larger clusters.

To characterize structural changes in polymers, we compute the gyration tensor $\mathbf{S}$~\cite{Gyration2013} 
\begin{equation}
    \mathbf{S} = \frac{1}{N_m} \sum_{i=1}^{N_m} (\mathbf{r}_i - \mathbf{r}_{\mathrm{cm}})(\mathbf{r}_i - \mathbf{r}_{\mathrm{cm}})^{\mathrm{T}},
\end{equation}
where $\mathbf{r}_i$ is the position of monomer $i$ within the polymer and $\mathbf{r}_{\mathrm{cm}}$ is the polymer's center of mass. The radius of gyration, quantifying the overall size of the polymer, is $R_g = \sqrt{\operatorname{tr}(S)}=\sqrt{\lambda_1+\lambda_2+\lambda_3}$, where $\lambda_1 \ge \lambda_2 \ge \lambda_3$ are the eigenvalues of $\mathbf{S}$. Polymer shape anisotropy is quantified by $b=\lambda_1-0.5(\lambda_2+\lambda_3)$, with larger $b$ indicating more elongated, rod-like conformations. As shown in Fig.~\ref{fig:wide}(b,c), polymer morphology depends strongly on both active stress and polymer concentration. Strong active stresses generate intense flow fields that stretch polymers, increasing both $R_g$ and $b$ and producing rod-like or cylindrical conformations. In particular, pullers with $\beta = 3$ significantly stretch polymers, yielding elongated shapes with $R_g/l^n_0 \approx 22.7$ and $b/l^n_0 \approx 700$. Additionally, increasing polymer concentration from $\phi_p=0.15$ to $\phi_p=1.0$ reduces elongation due to enhanced excluded-volume interactions that hinder stretching. Although polymers are well-extended at $\phi_p=0.15$, this concentration is too low to have a pronounced impact on regulating squirmer alignment. Polymers influence becomes more apparent only when the corresponding concentration reaches a certain threshold. Similarly, as shown in Fig.~\ref{fig:wide}(e,f), denser swimmer suspensions ($\phi_s = 0.15$) produce stronger flows that generally enhance polymer elongation. 

\begin{figure}[!htbp]
\includegraphics[width=0.9\columnwidth]{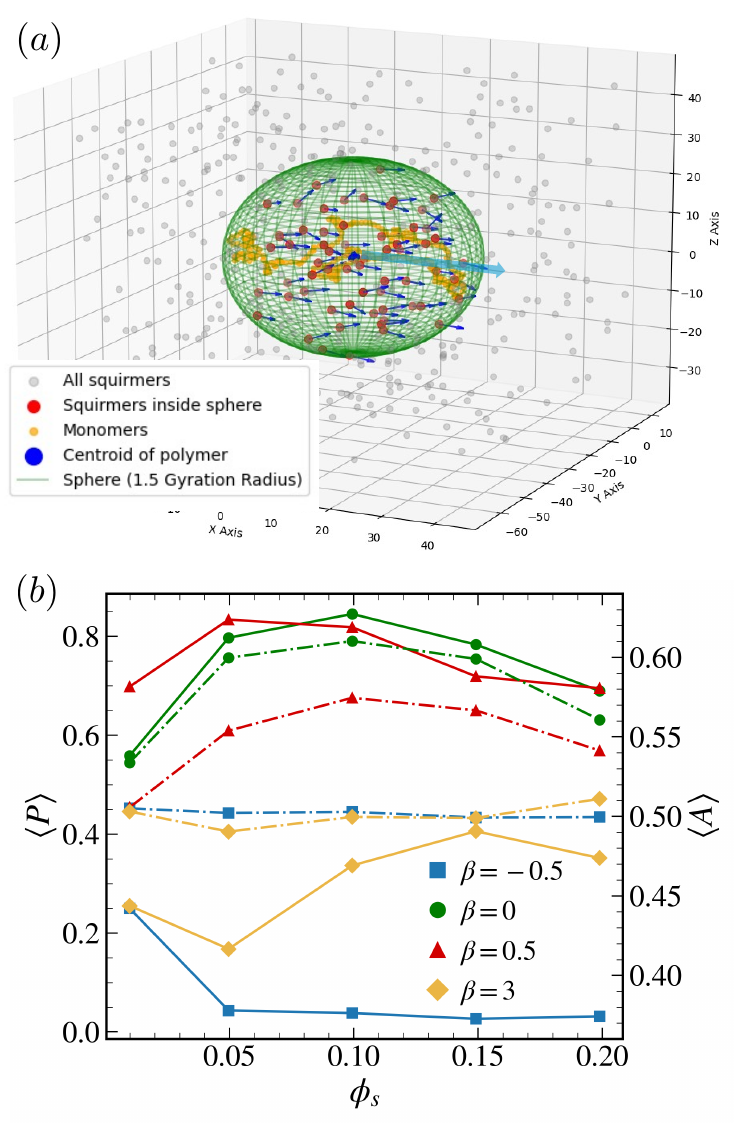}
    \caption{\label{fig:alignment} (a) Simplified simulation snapshot illustrating the alignment between a polymer and surrounding neutral swimmers within a spherical region of radius $1.5R_g$ at $\phi_s=0.1$ and $\phi_p=1.0$. The large light-blue arrow indicates the polymer’s principal stretching direction, while the small dark-blue arrows represent the swimmers’ orientations. (b) Relationship between the mean polarization $\langle P \rangle$ (solid lines) and mean alignment parameter $\langle A \rangle$ (dash--dot lines) as functions of squirmer concentration for various active stresses at $\phi_p=1.0$.}
\end{figure}

To verify the feedback–mechanism hypothesis, we introduce a correlation measure quantifying whether the alignment between polymer stretching directions and squirmer orientations contributes to the observed polarization enhancement. The alignment parameter is defined as
\begin{equation}
    A(t)=\frac{1}{N_p} \sum_{i=1}^{N_p} \left| \mathbf{u}_i \cdot \mathbf{u}^p_i \right|, 
\end{equation}
where $\mathbf{u}^p_i$ is the eigenvector corresponding to the largest eigenvalue of the gyration tensor of polymer $i$, representing its principal stretching direction, and $\mathbf{u}_i = \sum_{j=1}^{n_i} \mathbf{e}_j /|\sum_{j=1}^{n_i} \mathbf{e}_j|$ is the local polarization direction of $n_i$ squirmers within a spherical region of radius $1.5 R_g$ centered at the polymer. For two randomly oriented vectors in $d$-dimensional space, the probability density function of their relative angle $\theta$ is 
\begin{equation}
p(\theta)=\frac{\Gamma(d / 2)}{\sqrt{\pi} \Gamma((d-1) / 2)}(\sin \theta)^{d-2}, \quad \theta \in[0, \pi]
\end{equation}
where $\Gamma$ is the Gamma function, which generalizes the factorial function to continuous values \cite{cai2013distributions}. According to this distribution, the expected absolute inner product of two random vectors on a sphere is $\mathbb{E}[|\cos \theta|]=\frac{1}{2} \int_0^\pi|\cos \theta| \sin \theta d \theta= 1/2$. The alignment between a polymer and surrounding neutral swimmers is illustrated in Fig.~\ref{fig:alignment}(a), where excellent agreement between the polymer’s principal stretching direction and swimmer orientations can be observed, confirming the validity of the hydrodynamic feedback mechanism. As shown in Fig.~\ref{fig:alignment}(b), polarization and alignment exhibit strong coherence for weak active stresses, supporting the notion that polymer stretching and squirmer orientation are closely coupled. For neutral swimmers and pullers with $\beta = 0.5$, insufficient polymer stretching results in weak alignment at low swimmer concentrations. As the suspension becomes denser, intensified flow fields enhance polymer stretching and alignment, yielding peaks of $\langle A \rangle \approx 0.61$ for neutral swimmers and $\langle A \rangle \approx 0.58$ for pullers. At higher concentrations, increasingly complex hydrodynamic interactions disrupt this correlation, leading to a decline in alignment. For pushers and strong pullers, intense hydrodynamic disturbances make sustaining polymer--squirmer alignment more difficult, leading to diminished correlations. Similarly, as shown in Fig.~S8, the alignment parameter varies coherently with the polarization as the active stress changes, providing further support for the hydrodynamic feedback loop mechanism. 

Recent experiments on dilute sperm suspensions have shown that both cluster size and orientational alignment of swimmers increase with fluid viscoelasticity \cite{Tung2017sperm}. Similar behavior has been reported in 2D hydrodynamic simulations of rod-like pushers, where viscoelasticity enhances aggregation by generating strong polymer stresses in the interstitial gaps between aligned swimmers \cite{Li2016Collective}. In contrast, the mechanism in our system is different. By mapping the shear generated by squirmer motion to that produced by oscillatory walls, one can employ traditional oscillatory shear rheometry to measure the dynamic moduli of polymers under swimmer-induced flows \cite{qi2025protocol}. Our measurements indicate that the flexible polymer solutions considered here are predominantly fluid-like. Therefore, the observed polarization enhancement arises from the polymer-swimmer hydrodynamic feedback mechanism, rather than from bulk fluid elasticity.

We next examine whether polymer-squirmer or polymer-polymer entanglements contribute to the observed order enhancement. The mean number of monomers in contact with a squirmer is $2 \sim 6$, corresponding to a contact length of $1 \sim 5 l^n_{0}$, which is much smaller than the squirmer perimeter (see End Matter). This rules out squirmer-polymer entanglement as the cause of polarization enhancement. To assess polymer-polymer entanglement, we use the Z1+ algorithm~\cite{Martin2023Z1+} to compute the mean number of entanglement-like points per chain $\langle Z \rangle$ (see End Matter). As shown in Fig.~\ref{fig:P_Z}, $\langle Z \rangle$ increases linearly with squirmer concentration $\phi_{s}$, reflecting enhanced interchain engagement driven by squirmer activity and the reduced effective free volume. However, this trend does not correlate with the behavior of polarization, indicating that polymer--polymer entanglements are likewise not responsible for the ordering.

Our simulations reveal that polymer deformation plays a central role in controlling microswimmer collective behavior through a hydrodynamic feedback mechanism. This effect is reminiscent of the dynamics of microswimmers in nematic liquid crystals (LCs), where active stresses destabilize nematic order, driving a transition from a uniform state to a continuously twisting, cholesteric-like structure. In that case, swimmer trajectories follow the LC director field and become helical in the cholesteric state \cite{gautam2024microswimmers}. In our system, polymers act as soft, deformable confinements that regularize swimmer motion via the hydrodynamic feedback mechanism. These findings suggest a pathway to externally control microswimmer order using polymers. For instance, electrically responsive polymers could enable modulation of swimmer dynamics via applied electric fields.

\textit{Acknowledgments}\textemdash{}This work has been supported by National Natural Science Foundation of China grants 12304257 and 12574237. 

\textit{Data availability}\textemdash{}The data that support the findings of this article are not publicly available. The data are available from the authors upon reasonable request.


\bibliography{manuscript}


\section{End Matter}

\subsection{Squirmer-polymer entanglement}

\begin{figure}[!htbp]
    \centering
    \includegraphics[width=0.95\columnwidth]{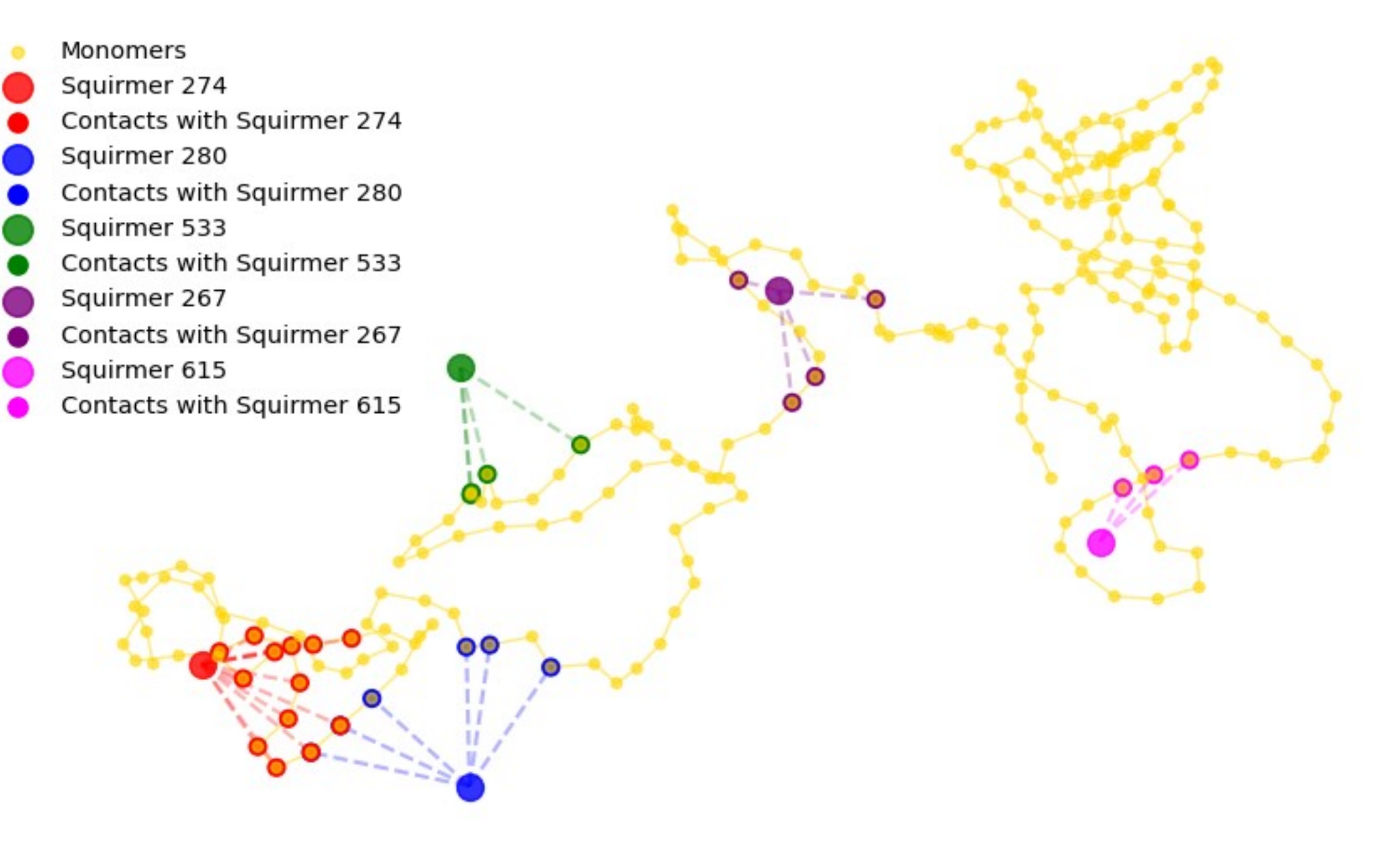}
    \caption{Schematic illustration of a polymer interacting with multiple squirmers at $\phi_s=0.15$ and $\phi_p=1.0$. A monomer is considered in contact with a squirmer when their center-to-center distances are less than $d=4.0$. For visualization, monomers within this defined contact distance are colored to match the corresponding squirmer's color.}
    \label{fig:sketch_squ_polu_contact}
\end{figure}

\begin{figure}[!htbp]
    \centering
    \includegraphics[width=0.95\columnwidth]{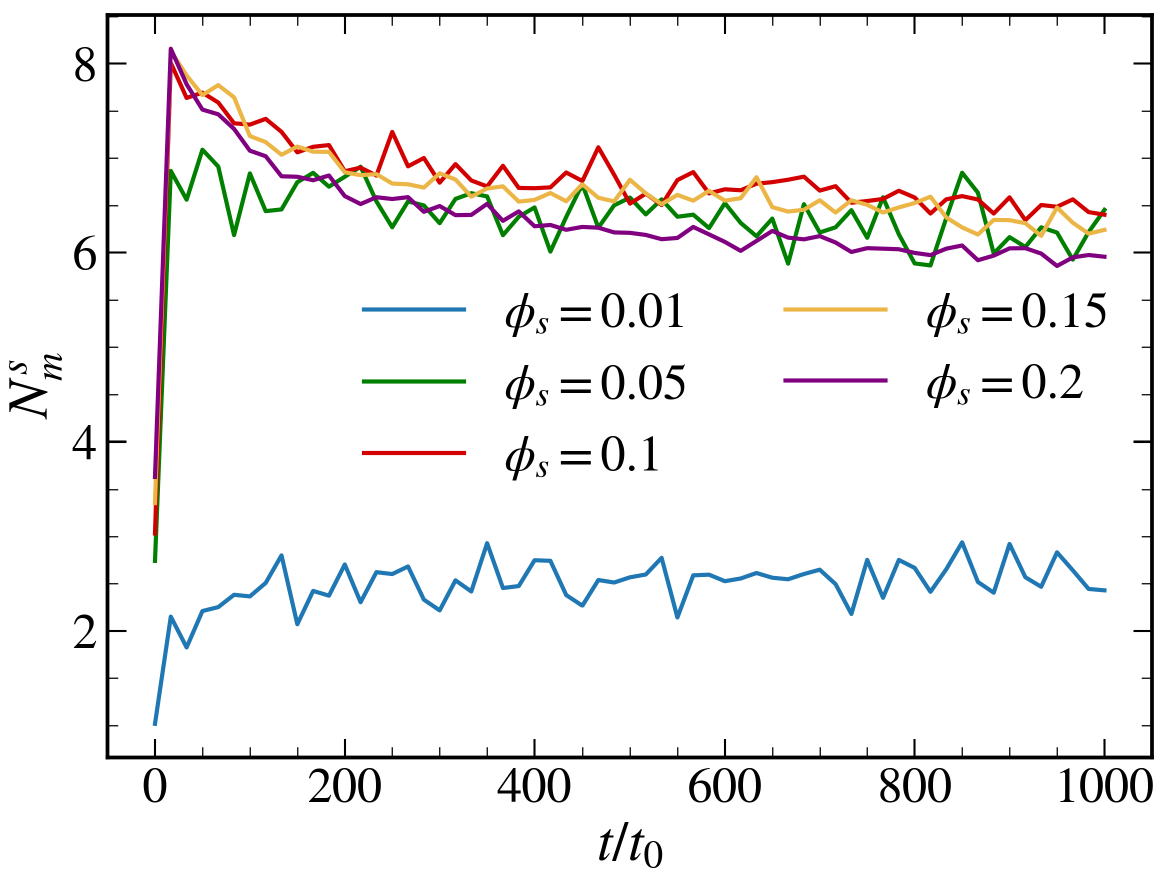}
    \caption{Average number of monomers in contact with a neutral swimmer as a function of time for various squirmer concentrations $\phi_s$ at $\phi_p=1.0$.}
    \label{fig:squ_poly_contact}
\end{figure}

To investigate whether entanglement or direct interactions between polymers and squirmers contribute to observable polarization changes, we quantified the average number of monomers in contact with a squirmer. A contact was defined as occurring when the center-to-center distance between a monomer and a squirmer satisfied $d/l^n_0 \leq 4$ (Fig.~\ref{fig:sketch_squ_polu_contact}). This threshold was chosen to capture close-range interactions, since it exceeds the sum of the squirmer radius ($R/l^n_0 = 3$) and the monomer radius ($r/l^n_0 = 0.2$), while still lying within the effective range of the soft-sphere potential (with a surface--surface cutoff distance of $r^{mm}_{c}/l^n_0 = 1.2$, corresponding to a center-to-center distance of $4.4$). As shown in Fig.~\ref{fig:squ_poly_contact}, the average number of monomers in contact with a squirmer, $N^{s}_{m}$, remains consistently below 9. This low contact frequency suggests that direct interactions or possible ``entanglements'' between polymers and squirmers are minimal. Therefore, these effects are not the primary drivers of the significant polarization changes observed in the system.

\subsection{Polymer-polymer entanglement}

\begin{figure}[!htbp]
    \centering
    \includegraphics[width=0.95\columnwidth]{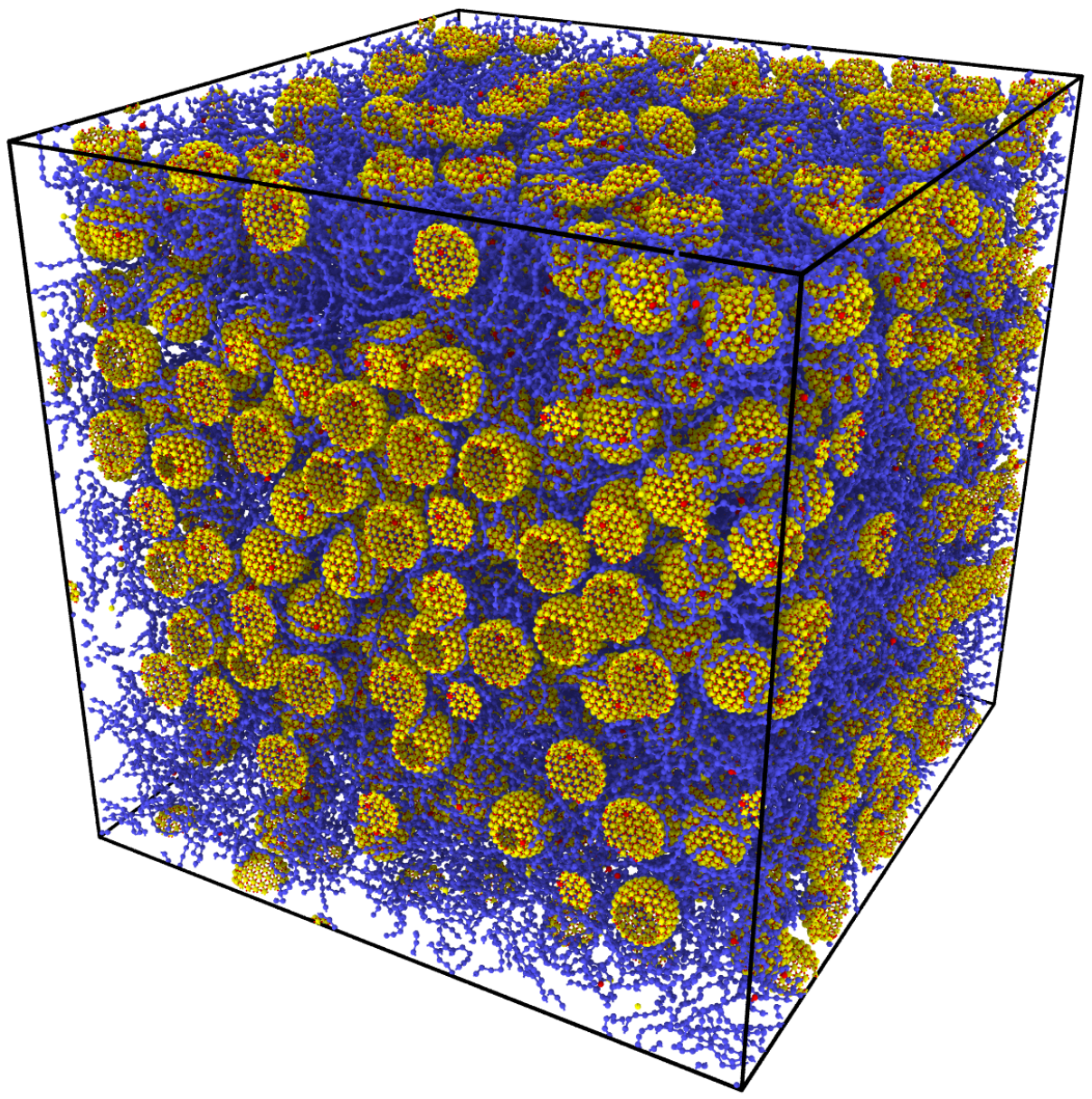}
    \caption{Schematic of the squirmers and polymers model used for Z1+ calculations at $\phi_s=0.15$ and $\phi_p=1.0$. The squirmer is represented as a spherical entity composed of multiple dumbbells connected head-to-tail. Polymer chains are shown with their starting ends in red, terminating ends in yellow, and intermediate monomers in blue. Each dumbbell in the model corresponds to two linked endpoints.}
    \label{fig:entangle}
\end{figure}

Building upon the observation that squirmers induce significant stretching of polymer chains, manifested as a large radius of gyration, we then investigated the resulting polymer entanglement. To quantify this, we employed the Z1+ software \cite{Martin2023Z1+}, which utilizes shortest multiple disconnected path analysis to determine the number of geometric entanglement points, denoted by $Z$. As depicted in Fig.~\ref{fig:entangle}, for this specific entanglement analysis, the squirmer model was conceptualized as a spherical entity constructed from interconnected dumbbells, which interacts with the polymer network. 

\begin{figure}[!htbp]
    \includegraphics[width=0.95\columnwidth]{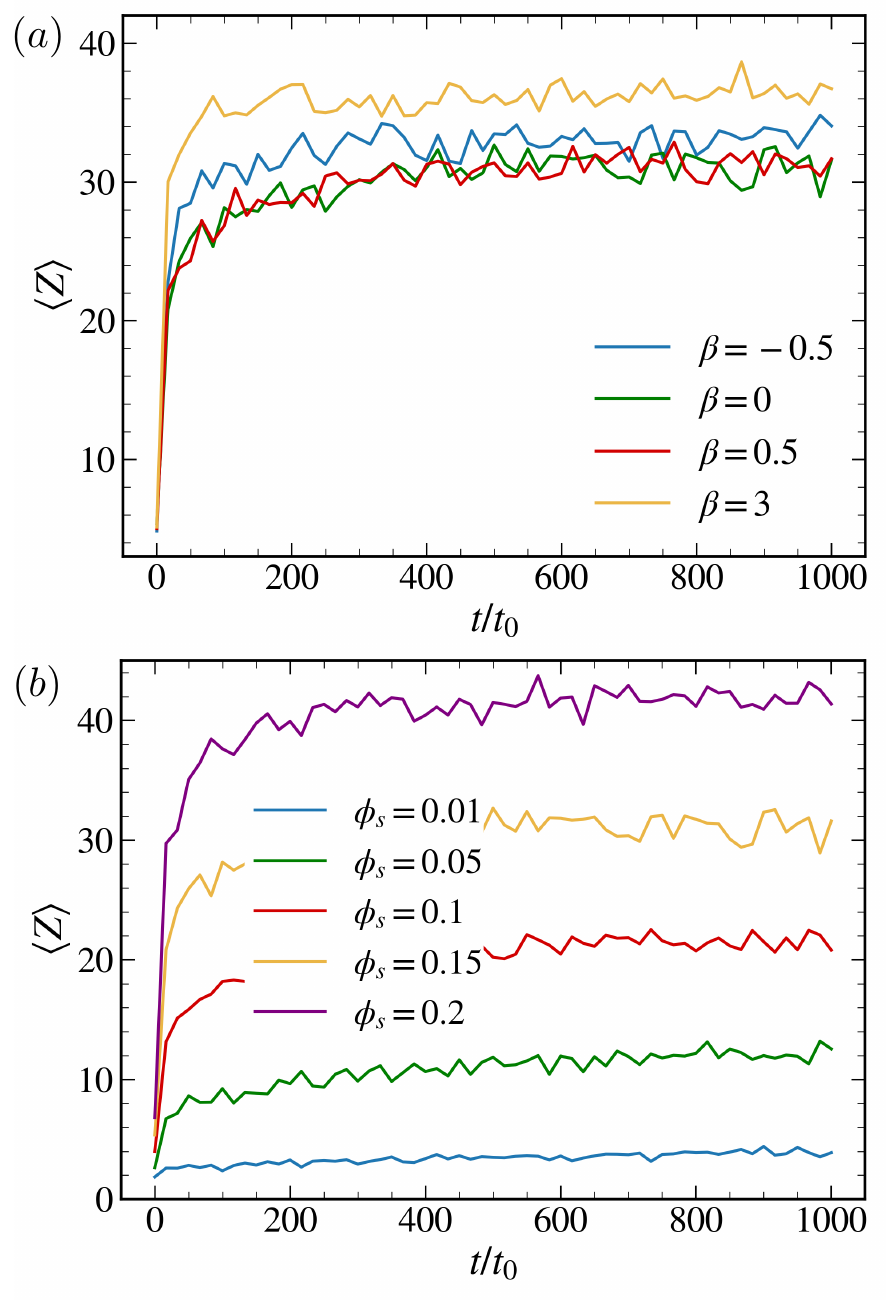}
    \caption{(a)Dependence of average entanglement number $\langle Z \rangle$ on the active stress parameter $\beta$, with squirmer concentration $\phi_s=0.15$ and polymer concentration $\phi_p=1.0$. (b) Average entanglement number $\langle Z \rangle$ as a function of squirmer concentration $\phi_s$ for neutral swimmers at fixed polymer concentration $\phi_p=1.0$.}
    \label{fig:Z}

\end{figure}

As shown in Fig.~\ref{fig:Z}, the time series of $Z$ rapidly converge to constant values. These equilibrium values are then averaged to obtain $\langle Z \rangle$ as a function of $\phi_s$. 

\begin{figure}[!htbp]
    \includegraphics[width=0.95\columnwidth]{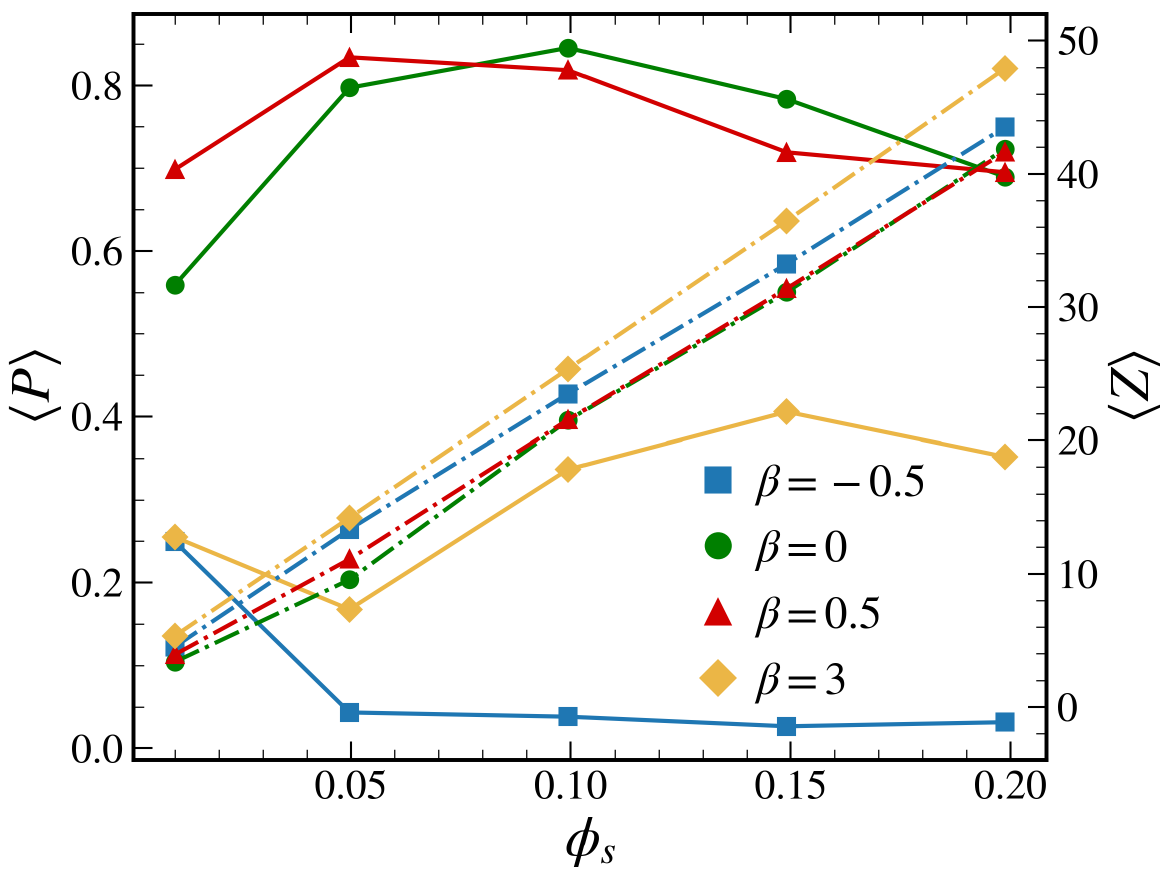}
    \caption{Relationship between the mean polarization $\langle P \rangle$ (solid lines) and mean number of entanglement points $\langle Z \rangle$ (dash–dot lines) as functions of squirmer concentration for various active stresses at $\phi_p=1.0$.}
    \label{fig:P_Z}
\end{figure}

As shown in Fig.~\ref{fig:P_Z}, $\langle Z \rangle$ generally increases with squirmer concentration $\phi_s$, signifying enhanced inter-chain engagement due to squirmer activity. Intriguingly, $\langle Z \rangle$ also exhibits a strong dependence on active stress. This hierarchy can be attributed to the interplay between the strength of the squirmer-generated flow field, the resultant polymer stretching ($R_g$), and the degree of induced polymer alignment. Specifically, pullers ($\beta=3$) generate intense flows that maximally elongate chains, increasing their $R_g$ and effective volume, thereby promoting the most frequent inter-chain crossings coupled with low flow-induced alignment. Conversely, neutral swimmer or puller ($\beta=0.5$) induce higher polymer alignment and less chain extension, resulting in comparatively lower values of $\langle Z \rangle$. Intermediate behavior is observed for pushers ($\beta=-0.5$), where moderate stretching combines with lower alignment. Ultimately, our findings show a clear decoupling between polymer entanglement and system polarization.

\end{document}